\documentclass[conference,9pt]{IEEEtran}
\usepackage{graphicx}
\usepackage{booktabs}
\usepackage{listings}
\usepackage{amsmath}
\usepackage{float}
\usepackage{amssymb}
\usepackage{cite}
\usepackage{xcolor}
\usepackage[hidelinks,pdfauthor={},pdfcreator={},pdfsubject={}]{hyperref}

\title{Verify the Verifier: Specification-Grounded LLM Cross-Checking for Bugs That Escape UVM Scoreboards}

\author{\IEEEauthorblockN{Ana E. Sanchez Villalobos, Michael D. Quinn, Aakash Tyagi, Minh D. Luu, David Kebo Houngninou}
\IEEEauthorblockA{Texas A\&M University, College Station, TX, USA\\
\{anaeugenia.sv, m.d.quinn, tyagi, minh.luu, davidkebo\}@tamu.edu}}

\begin{document}
\raggedbottom
\pagestyle{plain}
\maketitle 
\begin{abstract}
A passing UVM regression shows the design agrees with the reference model on what the scoreboard checks, not that it meets the specification. Bugs that pass every test yet violate the specification, termed scoreboard escapes, survive mature testbenches. We present Verify the Verifier, to our knowledge the first framework whose LLM-based Anomaly Cross-Checker examines a passing regression's traces against the specification. Of 564 bugs inserted into four OpenTitan IPs, 68 are candidate scoreboard escapes: non-equivalent, exercised by the tests, yet passing. The framework confirmed 41 as scoreboard escapes with assertions failing on the buggy design and proven on the original.

\end{abstract}


\section{Introduction}
UVM is the industry-standard verification methodology for constructing and reusing testbench components in the functional verification of digital designs. 
Late in the verification cycle, once the testbench is stable and the design has passed basic health checks, teams run large regressions, and a fully passing regression is one of the exit criteria the design must meet before it advances to the next milestone, such as RTL freeze or tapeout.
A bug that escapes this regression is therefore carried into later stages of the design flow, where it is more expensive to find and fix.

A passing regression, however, guarantees less than its role as an exit criterion implies. A UVM scoreboard compares the transactions observed at the design’s outputs with those a reference model predicts for the stimulus each test applies, so a passing regression shows only that the two agreed on the behavior the tests exercised and the scoreboard compared. Functional coverage reports which behaviors the test reached, but only among those someone chose to model: a behavior without a covergroup is neither measured nor checked. Neither certifies that the design complies with its specification. 

We define a scoreboard escape as a bug that passes the full regression yet still violates a specification requirement. Such bugs exist even in mature, public environments. The scoreboard of the OpenTitan PWM controller, for example, never measures the phase delay, skips pulses with 0\% and 100\% duty cycles, and silently discards any pulse whose period differs from the expected one, so a bug that changes the period is never compared and the regression still reports a PASS.

Recent LLM-based work on RTL debugging and repair starts from a failure. Diagnosis methods such as VCDiag \cite{vcdiag} analyze a failing simulation, and repair frameworks such as MEIC \cite{meic} and UVLLM \cite{uvllm} iterate until the tests pass, so a passing regression is treated as the end point, not examined for undetected scoreboard escapes. LLMs have also been used to generate stimulus \cite{zhang2025llm4dvusinglargelanguage} and assertions \cite{wu2025spec2assertionautomaticprertlassertion}, but to our knowledge none inspects the traces of the regression that already passes. Finding scoreboard escapes therefore raises two challenges. The first is observability: an escape is usually visible only in the behavior of internal signals, which the scoreboard does not see. The second is trust: a failing LLM-generated assertion is not, by itself, a detection, because it may reveal either a bug in the design or an error in the assertion itself, so it must be qualified independently (Section~\ref{sec:problem}).

\begin{figure*}[t]
 \centering
 \fbox{\includegraphics[width=0.7\textwidth]{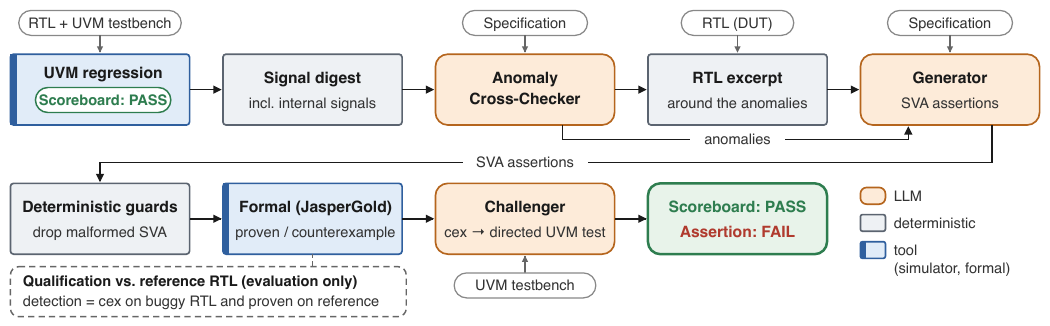}}
 \caption{Overview of Verify the Verifier. A regression that passes the
 scoreboard is inspected by the Anomaly Cross-Checker; the resulting
 assertions are checked with formal verification, and each counterexample
 becomes a directed UVM test that passes the scoreboard but fails the
 assertion.}
 \label{fig:overview}
\end{figure*}

To address these two challenges, we propose Verify the Verifier, a framework that treats a passing regression as evidence to be inspected rather than a verdict to be trusted (Fig.~\ref{fig:overview}). For observability, its core component is the Anomaly Cross-Checker (ACC), an LLM that receives the specification and a waveform trace of the simulation, including internal signals, and flags behavior that may contradict the specification even though the scoreboard reported a PASS. The ACC never sees the RTL, so its findings come from what the design did, not from how it was written. An Assertion Generator LLM (the Generator) then turns these anomalies into SystemVerilog assertions, using the specification and an excerpt of the RTL that drives the flagged signals. For trust, deterministic guards discard malformed assertions, and formal verification either proves each assertion or falsifies it with a counterexample. For each counterexample, an Adversarial Challenger LLM writes a directed test with the testbench's own UVM sequences, so that a finding becomes a reproducible test that passes the scoreboard but fails the assertion. 

To our knowledge, this is \textbf{the first framework to use an LLM to cross-check passing regression traces against the specification and formally qualify each scoreboard escape}. This work makes the following contributions. First, we formulate the problem of scoreboard escapes and define a criterion for qualified detection: an assertion counts only if it fails with a counterexample on the bug-inserted design and is proven, non-vacuously, on the reference (Section~\ref{sec:problem}). Second, we present the Anomaly Cross-Checker, which compares a signal digest against the specification without access to the RTL, so that its findings come from the design’s behavior rather than its implementation (Section~\ref{sec:anomaly}). Third, we present an assertion pipeline that combines an RTL excerpt around the flagged signals, deterministic syntax guards, formal verification, and directed UVM tests that target the counterexamples (Sections~\ref{sec:generation} -~\ref{sec:challenger}). Fourth, a benchmark of 68 non-equivalent scoreboard escapes in four OpenTitan IPs,
obtained by adapting BugGen \cite{buggen} to keep the mutants that pass the regression (Section~\ref{sec:setup}). Finally, we evaluate the framework on four OpenTitan IPs with their native UVM environments (Section~\ref{sec:results}). The ACC reported the effect of the injected bug for 55 of 68 candidate escapes, and for 41 of these, the framework produced a qualified detection. On the designs without injected bugs, the ACC reported only places where OpenTitan's documentation contradicts its own RTL, which is itself a finding about the reference documentation.

\section{Background and Related Work}
\label{sec:related}

A UVM scoreboard compares observed DUT behavior against expected results derived from the reference model. However, its checks are limited to the behaviors captured by the verification environment and the properties explicitly implemented in the scoreboard. A test passes when none of these comparisons fails, so its verdict is bounded by what the reference model predicts and what the monitor captures. A behavior that does not follow the specification but still produces an output the scoreboard accepts goes undetected.

LLM-based work on RTL debugging and repair starts from a failure. VCDiag \cite{vcdiag} triages failing simulations from their waveforms, while repair frameworks such as MEIC \cite{meic} and UVLLM \cite{uvllm} iterate between an LLM and the simulator until the tests pass. In all these flows, a passing regression is an end point, even though UVLLM reports that repairs which pass the given test cases can still overfit to them \cite{uvllm}. We instead start from a regression in which every test passes and use its simulation traces to look for behavior that contradicts the specification, which later in the design flow can become a serious bug.

AssertLLM \cite{assertllm} generates assertions from the specification document and evaluates them by formal property verification against a golden RTL, counting an assertion as correct if it holds on the golden design and incorrect if it fails. We use the same kind of check for a different purpose: to decide whether an assertion that fails on a design is evidence of a bug or an error in the assertion. Our assertions are not only derived from the specification, but also target the anomalous behavior observed in the simulation trace. Finally, BugGen \cite{buggen} uses LLM agents to insert functional bugs into the RTL and retains only the mutants that the regression detects. We invert its acceptance rule to keep the mutants that pass the regression, which, after deduplication, formal equivalence checking and an activation check, yields a benchmark of candidate scoreboard escapes. BugGen is not part of our framework; we use it only to build this benchmark.

Mutation-based testbench qualification, available commercially in Certitude~\cite{certitude}, has long inspected passing regressions by injecting faults and checking whether any test detects them. It grades the testbench rather than repairing it, and it re-runs the regression for every fault. LLMs have also been applied to other parts of the verification flow: generating UVM testbenches~\cite{Ye_2025} and test stimulus~\cite{zhang2025llm4dvusinglargelanguage}, writing assertions from specifications before RTL exists~\cite{wu2025spec2assertionautomaticprertlassertion} or within UVM environments~\cite{systems12100390}, and benchmarking these tasks end to end~\cite{wan2025fixmeendtoendbenchmarkingllmaided}; Bennett and Eder survey machine learning for design verification more broadly~\cite{bennett2025reviewmachinelearningmicroelectronic}. None of these inspects the traces of a regression that already passes.

\section{Problem Formulation}
\label{sec:problem}

We define a \emph{scoreboard escape} as a bug for which every test in the regression passes, with no UVM$\_$ERROR or UVM$\_$FATAL reports, while the design still violates a specification requirement that the scoreboard does not check or checks incorrectly. Such escapes arise because the scoreboard checks only the behavior its monitors capture, and only under the conditions that the regression tests exercise.

A failing assertion is not, by itself, evidence of a bug. An assertion can fail without a bug when it is incorrect, for example when it bounds the timing of a requirement more tightly than the specification does. Conversely, it can pass without checking anything when it is vacuous, that is, when its precondition can never be reached. For every assertion, JasperGold returns one of three results: proven, counterexample (cex), or undetermined. A proof alone cannot reveal vacuity, so for each assertion we also check a cover property on its precondition; the proof counts as non-vacuous only if JasperGold reaches this cover. We call an assertion \emph{qualified} when three conditions hold: it has a cex on the mutant, it is proven on the reference RTL, and its precondition cover is reached on the reference RTL. An assertion with a cex on the reference RTL is a \emph{false positive}. Vacuous and undetermined assertions are not qualified. Simulation on the reference RTL is not enough evidence for this decision, because an incorrect assertion can stay silent under the stimulus of the regression tests. The reference RTL is used only because the bugs are injected, as an evaluation instrument, not as a framework requirement. 

Not every mutant is detectable by checking the design against its specification; we distinguish the following types of mutants. An \emph{equivalent mutant} behaves the same as the reference for every input sequence, so it is not a bug. A \emph{non-activated mutant} is non-equivalent, yet the regression stimulus never exercises the difference: every output in its trace matches the reference on every cycle, leaving nothing for a trace-based check to find. A mutant that changes only behavior the specification does not state, or that makes the design conform to it, violates no requirement. Only a mutant that violates a stated requirement can be detected by such a check, including ours. We discard any equivalent and non-activated mutants when building the benchmark (Section~\ref{sec:setup}). Activating a non-activated mutant would require new stimulus, which is outside the scope of this work: the framework inspects the regression tests the team already runs. Its reproduction step already writes directed tests (Section~\ref{sec:challenger}), so extending it to activate such mutants is a natural next step.

\section{Framework}

\subsection{Overview}
\label{sec:framework}

The framework uses only artifacts a verification team already has once its regression passes: the specification, the RTL, the UVM environment, and the regression tests with their coverage results. Fig.~\ref{fig:overview} shows how it processes them in five steps. Three of them use an LLM (the Anomaly Cross-Checker, the Generator, and the Adversarial Challenger); the other two are deterministic.

\begin{enumerate}
    \item \textbf{Observation.} The regression tests selected by coverage criteria (Section~\ref{sec:anomaly}) are re-run with waveform dumping, and each trace is converted into a \emph{signal digest}: a compact text form that keeps every value of every signal of the target module, including internal ones.
    \item \textbf{Cross-checking.} The Anomaly Cross-Checker reads the signal digest and the specification and reports every behavior that may contradict the specification (Section~\ref{sec:anomaly}). The RTL is not used for this step.
    \item \textbf{Assertion generation.} From the anomaly report, the Generator writes SystemVerilog assertions from the specification and an excerpt of the RTL that drives the anomalous signals. Deterministic guards remove or repair malformed assertions (Section~\ref{sec:generation}).
    \item \textbf{Formal check.} The assertions are bound to the design, without modifying it, and checked with JasperGold. Each one is proven, refuted with a counterexample, or left undetermined (Section~\ref{sec:formal}).
    \item \textbf{Reproduction.} For each counterexample, the Adversarial Challenger writes a directed test with the testbench's own sequences. The test runs in the native environment, to show the assertion firing while the scoreboard still reports a pass (Section~\ref{sec:challenger}). If the existing sequences cannot reproduce a counterexample, the formal detection still stands; the test only adds a reproducible case.
\end{enumerate}

Two rules apply throughout the flow, one for checkers and one for stimulus. First, the behavior an assertion requires always comes from the specification, never from the implementation, so that the bug is not written into an assertion as correct behavior. The RTL enters only as an excerpt from which the Generator takes signal names and conditions, and the RTL query tool returns declarations, not logic. Second, stimulus may be derived from the design because a counterexample decides only where the test drives the design, while the verdict comes from an assertion grounded in the specification; as long as the stimulus stays within what the testbench can legally drive, a poor choice can miss a bug, but cannot create one.  

\subsection{Test Selection, Signal Digest Creation and Anomaly Cross-Checker}
\label{sec:anomaly}
The framework first uses per-test coverage from the regression tests to select the tests that the ACC will observe. A test is selected if it covers at least one coverage item that no other test covers, or if it exercises a region of the design whose coverage still has holes. Each selected test is then re-run at its original seed with waveform dumping enabled for the target module, so that the ACC sees exactly the behavior the passing regression produced. The resulting waveform is converted into a signal digest, a compact text form of the trace described below, and one digest per test is given to the ACC together with the specification.

The signal digest covers every signal of the target module, including internal ones, sampled at every rising clock edge. Instead of writing the value of each signal at every cycle, it describes how the signal behaves over a range of cycles: it stays constant, counts by a fixed step, or repeats a short pattern. This makes the trace compact. PWM's smoke test alone includes 51,000 cycles of 389 signals, which would take millions of tokens as one row per cycle. The encoding is exact at cycle precision: before use, every signal digest is decoded and compared, signal by signal and cycle by cycle, against the values sampled from the simulation dump.

The signal digest, along with the specification, is given to the ACC: it never sees the RTL or the reference design. The ACC uses no retrieval (RAG) or external tools; the specification and the signal digest are placed directly in the model's context. Its task is to point out any behavior that may contradict the specification. For each anomaly it reports the signal, the test where it was found, the specification requirement, the observed behavior and the discrepancy between them. The prompt asks the ACC which behaviors in the trace do not match the specification, rather than whether the observed behavior is legal, because the latter starts from what the design did and invites the LLM to construct a reading of the specification under which that behavior is acceptable, so that an anomaly is explained away rather than reported. A deterministic filter then discards entries that the LLM itself marked as having no discrepancy. Because the LLM's output may vary from run to run, the ACC gets up to three attempts and keeps the first non-empty result. The results of different attempts are not merged, because their union would also merge the false anomalies of each run.

\subsection{Assertion Generation and Deterministic Guards}
\label{sec:generation}

The Generator writes assertions from four inputs: the specification, the anomaly report from the ACC, the RTL interface (names, widths, and direction, without any logic), and an RTL excerpt. It never receives any indication of which line was mutated. The RTL excerpt is built deterministically, without an LLM or a human involved: it takes the statements that drive each signal named in the anomaly report, then the statements that drive the signals read by those statements, up to two levels deep, crossing module boundaries only through port connections. The same rule is applied to every design and every mutant. All comments are removed before the excerpt reaches the Generator.

What an assertion requires always comes from the specification. The RTL excerpt given never supplies the required behavior; it only provides exact signal names, their structural relations, and enabling conditions. An enabling condition is an extra condition in the antecedent that limits when a requirement applies. An enabling condition is only used if it does not contradict the specification, and every assertion records what it took from the excerpt and why it does not contradict the specification. The Generator can also call \texttt{grep\_rtl}, a tool that returns the exact declaration of a signal but never the logic behind it, so the LLM cannot read the design's behavior and turn it into the reference for its assertions. 

Recurring LLM mistakes are caught by deterministic code rather than left to the prompt, so they never reach formal verification. A signal validator discards assertions that use identifiers that are not RTL signals or locally declared names. A binder then wraps all remaining assertions into a single checker module that is attached to the DUT with a SystemVerilog \texttt{bind}, without modifying the RTL. The same checker is used in formal verification and in simulation. This binder also drops or fixes any construct that would break the compilation, for example non-constant repetitions or local names that can shadow DUT signals. Every guard acts only on the offending assertion, so the remaining assertions are still compiled and checked.

\subsection{Formal Verification}
\label{sec:formal}

The checker module is checked with JasperGold on the DUT under a time limit on the proof run, ensuring that a property whose proof does not converge cannot stall the flow. Each assertion is reported as proven, counterexample (cex), or undetermined. 
For each assertion, JasperGold also checks a cover property on its antecedent, which identifies vacuous proofs. A proven assertion is true in every reachable state of the design, so the framework takes no further action on it, and an undetermined one is set aside. For an assertion that fails, JasperGold produces a counterexample: a cycle-by-cycle trace of input values that drives the design from reset to a state that violates the assertion. This trace is the input to the Adversarial Challenger (Section~\ref{sec:challenger}), while whether a cex counts as a detection in our evaluation is decided by the qualification criterion shown in Section~\ref{sec:problem}.

\subsection{Adversarial Challenger and Simulation Confirmation}
\label{sec:challenger}
The Adversarial Challenger LLM turns each cex into a directed UVM test. It receives the violated assertion, the cex trace produced by JasperGold, the testbench's own context (its base virtual sequence, register-access tasks, and environment package), and the RTL declarations, but not its logic. Its task is to reproduce the trace, not to search for a different attack. Its output is a virtual sequence that extends the environment's base sequence and drives only the inputs the testbench can drive, through the testbench's own UVM register-access tasks, so the scoreboard and monitor stay consistent with the stimulus. Each sequence is compiled against the environment and discarded if it fails, without affecting other sequences.

Each accepted sequence runs as its own test, starting from reset, in the original UVM environment with the assertion checker bound to the DUT. An assertion failure is reported by the simulator as an error but not as a UVM error, so the scoreboard result is not affected. The framework counts assertion failures and records the scoreboard verdict separately. A finding is a test in which an assertion fails, together with the scoreboard verdict observed on that same test.

\section{Experimental Setup}
\label{sec:setup}

\subsection{Designs and Verification Environments}

We evaluate our framework on four IPs from OpenTitan \cite{opentitan} (tag \texttt{earlgrey\_1.0.0}): PWM (Pulse-Width Modulator), PATTGEN (Pattern Generator), SPI Host and CSRNG (Cryptographically Secure Random Number Generator). All four are used with OpenTitan's own UVM environments, which we used without any modification. Their regression tests run through OpenTitan's \texttt{dvsim} flow on Cadence Xcelium. The regression of each IP is a set of functional tests from its test plan (4 for PWM, 5 for PATTGEN, 12 for SPI Host and 5 for CSRNG). The shared register, alert and bus-integrity tests are excluded because they do not exercise the block's function. One SPI Host test (\texttt{status\_stall}) is also excluded, because it fails on the unmodified reference at the fixed seed. Using a third-party environment is intentional: since we did not write any part of the verification environment or the design, no escape can be attributed to a checker we weakened. All four scoreboards have blind spots in their transaction-level checking. The PWM scoreboard compares only pulses whose period matches its prediction, silently skipping the rest, and never measures the phase delay. The PATTGEN scoreboard compares only the serialized data bits and checks neither the clock divider's timing nor the idle output levels. The SPI Host scoreboard compares only the transmitted and received bytes; chip-select timing parameters are never checked. The CSRNG scoreboard compares the generated bits and the command status, but not the alert outputs: its error test turns off the fatal-alert check, and only field-encoding errors are tested to raise the recoverable alert.

The specification used is OpenTitan's own documentation of each IP (overview, theory of operation, programmer's guide, register and interface descriptions), concatenated in a single file. PWM consisted of 655 lines, 424 lines for PATTGEN, 2076 for SPI Host and 1602 for CSRNG. We add no clauses, annotations nor corrections; where the documentation and the RTL disagree, the documentation is kept as is. The framework targets the module that holds each IP's logic: \texttt{pwm\_core}, \texttt{pattgen\_chan}, \texttt{spi\_host\_fsm} and \texttt{csrng\_core}, whose ports carry the register configuration as packed structs, which the signal digest decodes into named fields.

\subsection{A Benchmark of Scoreboard Escapes}

No public benchmark of scoreboard escapes exists: bug trackers, such as OpenTitan's, record bugs once they are found, not the bugs a passing regression missed. We therefore build one using BugGen \cite{buggen}. This tool inserts one LLM-written, single-line mutation per run into a functional region of the RTL, using GPT-4o-mini. We invert its acceptance rule: a mutant that compiles and passes the complete regression is kept as an escape candidate, and one that fails any test is set aside, because the regression already detects it. To validate these bugs, candidates then pass three filters. The first filter is \textbf{Deduplication}: identical mutants, and mutants that differ only by an operator equivalent at the given width (e.g., \texttt{\textasciitilde} vs.\ \texttt{!} on a one-bit signal), are merged. The second is \textbf{Equivalence}: we place the reference and the mutant side by side in JasperGold, driven by the same clock, reset and inputs, and ask the tool to prove that their outputs are equal on every cycle, with the inputs unconstrained except for legal register writes in PWM and the design's own embedded assumptions in CSRNG. If JasperGold produces a counterexample, the mutant is confirmed as non-equivalent; if it proves that the outputs are equivalent under all reachable states, the mutant is classified as equivalent; the undetermined ones are excluded. Finally, the \textbf{Activation} filter simulates the reference and the mutant under the regression and compares their traces signal by signal. A real mutant whose outputs are identical to the reference's is never exercised by the regression, so no trace-based method can observe it. Table~\ref{tab:funnel} shows how many bugs remain after each step: of the 241 that pass every regression test, 68 remain after the three filters and are evaluated end to end. Besides the three filters, some mutants are excluded as \emph{misc}: for example because they do not elaborate in JasperGold, leave an output undriven, reside in a submodule the checker cannot be bound to,  or produce traces that exceed the LLM’s context window. We call these 68 mutants candidate escapes: they pass every test and change the design's behavior, but only a qualified assertion establishes that the change violates a stated requirement. The 41 mutants the framework detects (Table~\ref{tab:funnel}) are therefore confirmed escapes. Some of the other 27 may violate no stated requirement, so the true detection rate is at least 41 of 68. 

\begin{table}[!ht]
\centering\footnotesize
\caption{Benchmark construction, from inserted bugs to the scoreboard escapes
confirmed by the framework.}
\label{tab:funnel}
\setlength{\tabcolsep}{4pt}
\begin{tabular}{@{}lrrrrr@{}}
\toprule
 & PWM & PATTGEN & SPI Host & CSRNG & Total \\
\midrule
Bugs inserted (at least)        & 62 & 124 & 60 & 318 & 564 \\
Pass all regression tests       & 46 & 84  & 50 & 61  & 241 \\
\quad $-$ Duplicate             & 3  & 33  & 10 & 4   & 50  \\
\quad $-$ Equivalent            & 4  & 7   & 14 & 9   & 34  \\
\quad $-$ Undetermined          & 3  & 3   & 0  & 19  & 25  \\
\quad $-$ Not activated         & 2  & 22  & 10 & 14  & 48  \\
\quad $-$ Misc                  & 10 & 3   & 0  & 3   & 16  \\
\midrule
Evaluated (candidate escapes)   & 24 & 16  & 16 & 12  & 68  \\
Confirmed by the framework      & 15 & 8   & 14 & 4   & 41  \\
\bottomrule
\end{tabular}
\end{table}

\subsection{Framework Configuration}

The three LLMs (Anomaly Cross-Checker, Generator and the Adversarial Challenger) use Claude Opus 5.5 with adaptive reasoning at high effort. Sampling is not deterministic, so each run is an independent sample. We evaluate each mutant with one framework run (two Generator runs for PWM20, mutant 20 of PWM; each mutant is named by its design and an index). If the ACC reports no anomaly, it is re-invoked up to twice; if it still reports none after three attempts, the Generator runs without an anomaly report. If the Generator produces no assertion, nothing reaches formal verification and the mutant counts as not detected. Formal verification uses Cadence JasperGold version 2024.09 with a 900 s proof time limit; an assertion not decided within that time is classified as undetermined.

\subsection{Evaluation Protocol and Metrics}

The evaluation protocol described in this section measures the framework but is not part of it. To decide whether a generated assertion is a genuine detection or a false alarm, we compare its behavior on the mutant against its behavior on the unmodified reference RTL. The reference RTL is used only after the framework has finished, and no stage of the framework reads it. In a real project, the bugs are unknown and no bug-free reference exists.

Each generated assertion is classified as \emph{qualified} or \emph{false positive} as defined in Section~\ref{sec:problem}, or as \emph{other} if it is undetermined, vacuous, or proven on both the mutant and the reference. A mutant is \emph{detected} if at least one of its assertions is qualified. 


\section{Results}
\label{sec:results}

We organize the results in two parts. The first evaluates the framework end to end on designs with known bugs: we report overall detection results (Section~\ref{sec:res-e2e}), then examine the two stages that determine them, the ACC and the Generator. The second evaluates the ACC on the unmodified reference designs, which contain no injected bug, to assess its specificity. All results refer to the 68 mutants evaluated end to end (Table~\ref{tab:funnel}).

\subsection{End-to-End Results}
\label{sec:res-e2e}

\begin{table}[!ht]
\centering\footnotesize
\caption{End-to-end results per design.}
\label{tab:e2e}
\setlength{\tabcolsep}{4pt}
\begin{tabular}{@{}lrrrrrr@{}}
\toprule
 & \multicolumn{3}{c}{Bugs} & \multicolumn{3}{c}{Generated assertions} \\
\cmidrule(lr){2-4}\cmidrule(l){5-7}
 & Tested & Confirmed & Reproduced & Caught & Wrong & Other \\
\midrule
PWM      & 24 & 15 & 15 & 28 & 27 & 6 \\
PATTGEN  & 16 & 8  & 8  & 11 & 31 & 0 \\
SPI Host & 16 & 14 & 12 & 25 & 19 & 3 \\
CSRNG    & 12 & 4  & 1  & 7  & 9  & 1 \\
\midrule
Total    & 68 & 41 & 36 & 71 & 86 & 10 \\
\bottomrule
\end{tabular}
\end{table}

Table~\ref{tab:e2e} shows the results of exercising the framework on the four designs. It produced a qualified detection for 41 of the 68 candidate scoreboard escapes: 15 of 24 in PWM, 8 of 16 in PATTGEN, 14 of 16 in SPI Host and 4 of 12 in CSRNG. In 36 of the 41, the Adversarial Challenger's directed test fired the qualified assertion on the mutant, the same test left it silent on the reference, and OpenTitan's scoreboard reported a pass. Each of these is therefore a bug that the native environment accepts and the framework both proves formally and reproduces in simulation.

The 27 candidate escapes that were not detected (68 evaluated, 41 detected) were missed at two different stages. The ACC observed the injected bug in 55 of the 68 mutants; in 14 of them the anomaly report was correct, but no generated assertion qualified. The 13 mutants the ACC did not observe are discussed in Section~\ref{sec:res-acc}. In other words, the framework correctly identifies the erroneous behavior in 81\% of the candidate escapes, and the remaining losses are split almost evenly between the ACC and the Generator. Per mutant, the framework took a median of 9.6 minutes, mostly simulation time, and about 620k input and 18k output tokens of LLM usage.

\subsection{Accuracy and Contribution of the Anomaly Cross-Checker}
\label{sec:res-acc}

\begin{figure}[!ht]
  \centering
  \includegraphics[width=0.62\columnwidth]{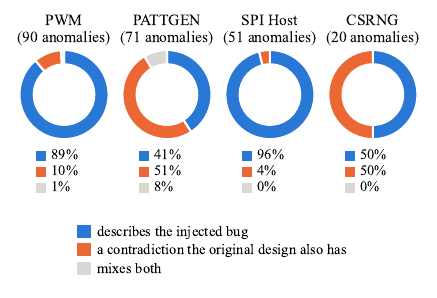}
  \caption{Anomalies reported by the ACC on the mutants, by what they describe.}
  \label{fig:acc}
\end{figure}

To evaluate the ACC, we compared each of its anomalies with the mutation after the run and gave it one of three labels: it describes the injected bug, it describes a contradiction that the unmodified reference also shows, or it mixes both (Fig.~\ref{fig:acc}). The ACC itself never sees the reference or the mutation. Of the 232 anomalies, 168 (72\%) describe the injected bug, and most of the remaining ones describe contradictions of the reference itself (Section~\ref{sec:res-golden}). The ACC observed the injected bug in 55 of the 68 mutants, always on its first attempt and without access to the RTL or to the location of the mutation.

The 13 mutants the ACC did not observe share one pattern: its report was taken by a contradiction that the unmodified reference also shows, such as the command-ready flag of CSRNG or the disabled channel of PATTGEN. The effect of the mutation was either not mentioned or mentioned only inside such an anomaly. In PWM17, for example, the ACC mentioned the mutation's effect (the beat counter is not cleared on a configuration write) only in passing, inside an anomaly whose main claim also holds on the reference, so it was counted as not observed. Several of these mutants remove a behavior rather than change it: CSRNG83 never raises the fatal alert, and an absence is harder to notice in a trace than a wrong value.

We also ran the same Generator and formal flow with only the specification and the RTL declarations, that is, without the signal digest, the anomaly report, and the RTL excerpt. Because this input does not depend on the mutant, the Generator ran once per design rather than once per mutant, and its assertions were checked against every mutant of that design. This specification-only baseline detected 11 of the 68 candidate escapes, compared with 41 for the full framework. Without a trace, the Generator writes plausible but generic properties, whereas an anomaly report directs it to the behavior that actually went wrong.

\subsection{Quality of the Generated Assertions}
\label{sec:res-gen}
Fewer than half of the assertions that reached formal verification qualified (71 of 167), and almost all others were false positives (86). In our evaluation, the formal check on the reference rejected every false positive automatically, so none of these was counted as a detection and each cost only one formal run. This includes several that fire on the mutant in simulation but stay silent on the reference under the same test, which a simulation-only check would have accepted. Most false positives are not hallucinated. In PATTGEN, PWM and CSRNG, they faithfully encode a documented requirement that the reference RTL does not implement, the same contradictions the ACC reports in Section~\ref{sec:res-golden}, such as the disabled PATTGEN channel. In SPI Host, most bound a wait state more tightly than the RTL counts it. We discuss how to reduce them in Section~\ref{sec:limitations}.

\subsection{The ACC on the Unmodified References}
\label{sec:res-golden}

Finding bugs is only half of the ACC's job; it must also stay silent on behavior that matches the specification. To check this, we ran the ACC with the same prompt, specification and attempt limit on the traces of the four unmodified designs, which contain no injected bug. It reported 9 anomalies (PWM 1, PATTGEN 6, SPI Host 1, CSRNG 1). None resembles any of the injected bugs, and each is a genuine contradiction between OpenTitan's documentation and its RTL, which we confirmed in the RTL source and in the trace:
\begin{itemize}
    \item \textbf{PWM}: the documentation describes an output register that delays the pin by one cycle; the RTL drives the output combinationally.
    \item \textbf{PATTGEN}: disabling a channel ``is equivalent to an FSM reset'' and returns its outputs to the inactive levels; the RTL keeps the channel active when it is disabled before the pattern completes (6 entries, one per output and channel).
    \item \textbf{SPI Host}: \texttt{clk\_cntr\_q} is documented as ``constantly held at zero'' in the idle states; the RTL holds its previous value and obtains the same timing through a separate enable term.
    \item \textbf{CSRNG}: for a generate command, the acknowledgment is documented to rise ``after all of the requested entropy is consumed''; the RTL raises it as soon as the bits are generated, before software reads them.
\end{itemize}

These contradictions reappear in the mutants' reports, and the PATTGEN contradiction explains all 31 PATTGEN false-positive assertions (Section~\ref{sec:res-gen}). 
On the CSRNG mutants, the ACC also flagged a second contradiction: the documentation keeps the command-ready flag low until the previous command is acknowledged, and the RTL does not. Every assertion built from it also fails on the reference. Without any bug, the ACC therefore raised no false alarm: every anomaly it reported on the references is a real contradiction between documentation and RTL. The false positives of Section~\ref{sec:res-gen} are assertions, written later by the Generator, and the formal check removes them.

\section{Discussion and Future Work}
\label{sec:limitations}

\paragraph{What the results show}
The framework finds escapes that a mature, native regression misses, and most of these detections depend on the ACC: a specification-only baseline detects 11 of the 68, against 41 with it. Its reports are also accurate. On the unmodified designs, every anomaly it reported was a genuine contradiction between documentation and RTL, and when it missed a bug, its report was dominated by such a contradiction  rather than an invented one.

\paragraph{What it delivers}
The framework reports behavior that contradicts the specification yet passes the testbench, which shows where the testbench does not fully check the specification. It also hardens the testbench: each confirmed finding leaves an assertion and, when available, a directed test that can be added to the regression, so the same behavior is caught if it reappears later for another reason. In a real project, no bug-free reference exists to qualify a failing assertion automatically, so each failing assertion goes to an engineer together with its anomaly and its counterexample, who decides whether it reveals a bug or is itself wrong.

\paragraph{Where and when to use it}
It is meant for a mature design and testbench, late in volume verification, when the bug rate is low, the regression passes and coverage is being closed. At that point it needs only what a team already has: the specification, the RTL, the UVM environment and passing regression traces. Unlike mutation-based testbench grading such as Certitude~\cite{certitude}, which reruns tests for every injected fault, it re-runs each selected passing test once, with waveform dumping, and can run whenever the regression passes.

\paragraph{Cost and return}
In our experiments, a run took a median of 9.6 minutes, mostly simulation, and about 620k input and 18k output tokens, which is small compared with the cost of a bug found after tapeout. The return falls when successive runs yield only anomalies already reviewed, such as known contradictions between documentation and RTL, and no new assertion survives the formal check. From then on, it is enough to rerun the framework only when the RTL or the testbench changes, since only a change can introduce new escapes. 

\paragraph{Where detections are lost, and next steps.}
Most missed detections come after the ACC: 86 of the 167 assertions failed on the reference RTL, and in 14 candidate escapes the ACC described the bug but no assertion qualified. These assertions fail for two reasons: they require tighter timing than the RTL implements, or they encode a documented requirement that the RTL does not follow. Our main next step is to give the Generator the design’s timing without exposing its logic.
The ACC missed 13 candidate escapes, mostly where documented behaviors the RTL does not follow drew its attention or a trace exceeded the context window; reading traces in parts addresses this. The Challenger did not reproduce 5 of the 41 detections, mainly because the testbench's own sequences cannot produce the counterexample's stimulus; letting it write new sequences would close this gap. 
Finally, the whole flow can be executed as an autonomous agent that rewrites failing assertions and generates new stimulus, while the formal check keeps every report trustworthy.
\paragraph{Threats to validity}
Each mutant was evaluated with one framework run, and each regression test ran at one seed, since our goal was to demonstrate the framework rather than grade these testbenches. OpenTitan's own configuration runs these tests with more seeds, so some of the 68 escapes might be caught by other seeds; such escapes reveal a gap in the stimulus rather than in the scoreboard’s checks, and the framework still finds them. OpenTitan is public, so the LLM could have seen its documentation and RTL during training. This does not explain our results: we injected the bugs ourselves, so the LLM cannot have seen them before, and it can find them only by reading the signal digest.

\section{Conclusion}

This work showed that an LLM can inspect the traces of a passing regression for the bugs it missed, when every finding it produces must survive a formal check. To our knowledge, Verify the Verifier is the first framework to do so: its Anomaly Cross-Checker reads the simulation traces a team already has, finds where the design contradicts its specification, and each finding becomes an assertion that formal verification must confirm. Across four OpenTitan IPs, the framework confirmed 41 scoreboard escapes, and the Anomaly Cross-Checker raised no false alarm on the unmodified designs. The weakest link is not finding the anomaly but writing the assertion that captures it, and that sets the initial direction of our future work. Closing this gap would turn a passing regression from an exit criterion into the starting point of a search for the bugs it missed.

\section*{Content Generated by AI}
Figure~\ref{fig:overview} was created with the assistance of a generative AI
tool and was reviewed and edited by the authors.

\nocite{*}
\bibliographystyle{IEEEtran}
\bibliography{refs}

\end{document}